# Deep Learning Interference Cancellation in Wireless Networks

Yiming Zhou, Ashkan Samiee, Tingyi Zhou, and Bahram Jalali

Electrical and Computer Engineering Department, UCLA, Los Angeles, CA 90095

**Abstract:**

With the crowding of the electromagnetic spectrum and the shrinking cell size in wireless networks, crosstalk between base stations and users is a major problem. Although hand-crafted functional blocks and coding schemes are proven effective to guarantee reliable data transfer, currently deep learning-based approaches have drawn increasing attention in the communication system modeling [1, 2]. In this paper, we propose a Neural Network (NN) based signal processing technique that works with traditional DSP algorithms to overcome the interference problem in realtime. This technique doesn't require any feedback protocol between the receiver and transmitter which makes it very suitable for low-latency and high data-rate applications such as autonomy and augmented reality. While there has been recent work on the use of Reinforcement Learning (RL) in the control layer to manage and control the interference, our approach is novel in the sense that it introduces a neural network for signal processing at baseband data rate and in the physical layer. We demonstrate this "Deep Interference Cancellation" technique using a convolutional LSTM autoencoder. When applied to QAM-OFDM modulated data, the network produces significant improvement in the symbol error rate (SER). We further discuss the hardware implementation including latency, power consumption, memory requirements, and chip area.

## 1. Introduction

With the rapid development of new wireless communication systems, the size of cell area is getting smaller and the problem of crosstalk between cells becomes more severe (Figure 1). To overcome this crosstalk, the common solution is that the receiver sends feedback to the transmitter which then responds by adjusting the transmission frequency. However, these feedback loops increase the latency and penalize real-time applications where low latency is required such as autonomy and augmented reality (AR). To overcome this issue, blind techniques are needed in which the receiver is able to remove the interference on its own, i.e. without having to communicate back to the transmitter.

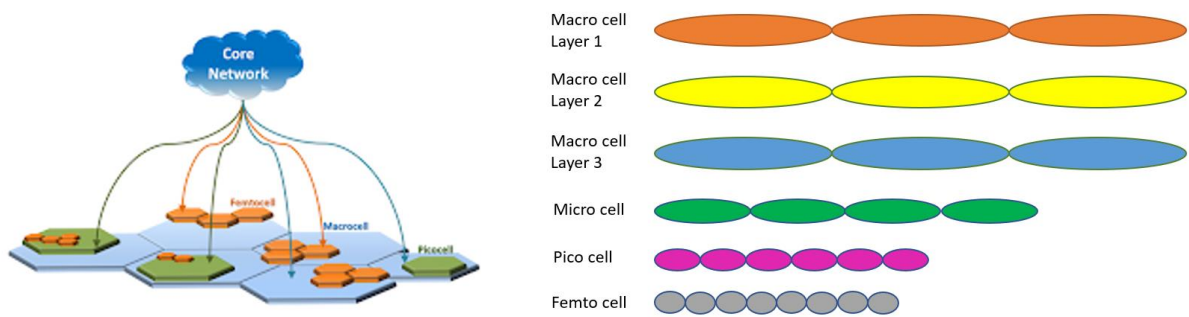

Figure 1: In order to offer more bandwidth to individual mobile users, the cell size has been getting smaller. As the cell size is reduced, inter-cell interference becomes a critical issue. Our approach of inserting AI signal processing blocks inside the receiver eliminates the crosstalk as well as mitigate other issues as described in the text. [3]

An example of interference in the LTE cellular communication system is shown in Figure 2. As the user equipment (UE) approaches the border of the two adjacent cells, there is a chance that there might be another user in the neighboring cell using the same frequency band for communication. This situation can cause interference to both users utilizing the same frequency band and dramatically reduce the data rate. To address this problem, the 3GPP consortium decided that the users at the same cell edge but belonging to different cells, use different frequency resources. Base stations that support this feature can generate interference information for each frequency resource (RB), and exchange the information with neighbor base stations through messages. Although this problem can fix the issue in low population areas, it does not work in crowded cells since the demand for empty bands increases, and the base has to use all the available bands for communication.

Another promising solution would be using phased array antennas at the receiver to implement spatial filtering to reject the interference. Although this technique can greatly reduce the power of the interference signal, it has shortcomings. The first issue is that phased array antennas are very susceptible to mismatches and since the fabrication process can easily add small mismatches, this can limit the amount of spatial filtering it could provide. In addition, the weights assigned to these antennas have to be adjusted adaptively to track the direction of the desired signal and keep rejecting the interference signal. These adjustments can increase the latency of the system since the link has to be probed and weights adjusted accordingly. Also, beamwidth is directly proportional to the number of available phased array antennas. As a result, there is a trade-off between the power, area, and beamwidth of the antenna.

There are several traditional digital signal processing (DSP) techniques to remove a narrow band interference from the desired wideband signal. However, they are not effective for the case of wideband interference, where the interference, can occupy the same bandwidth as the desired signal. In the OFDM/QAM modulation wireless systems, each resource element (RE) can have a discrete value from the QAM symbol space. If this is the case for both UE and the interferer, the combination of both of these two will result in a discrete sample space too. Using a Maximum Likelihood Estimation (MLE) approach for the classification of this new sample space requires a lot of computation and might not be practical. While conventional DSP techniques recover distorted signals by cascade operation of hand-crafted physics models to mitigate specific impairments sequentially, it is recognized that the emerging machine learning approaches based on neural networks may play a role in communication signal processing with reduced computation complexity and latency, as well as new capabilities in learning the complex behavior of signals [4].

In this paper, we propose a Neural Network (NN) architecture with the combination of convolutional layers and LSTM (long short-term memory) layers that can help the traditional DSP baseband chain to overcome this interference problem. This technique doesn't require any feedback protocol between the receiver and transmitter which makes it very suitable for low-latency and high data-rate applications.

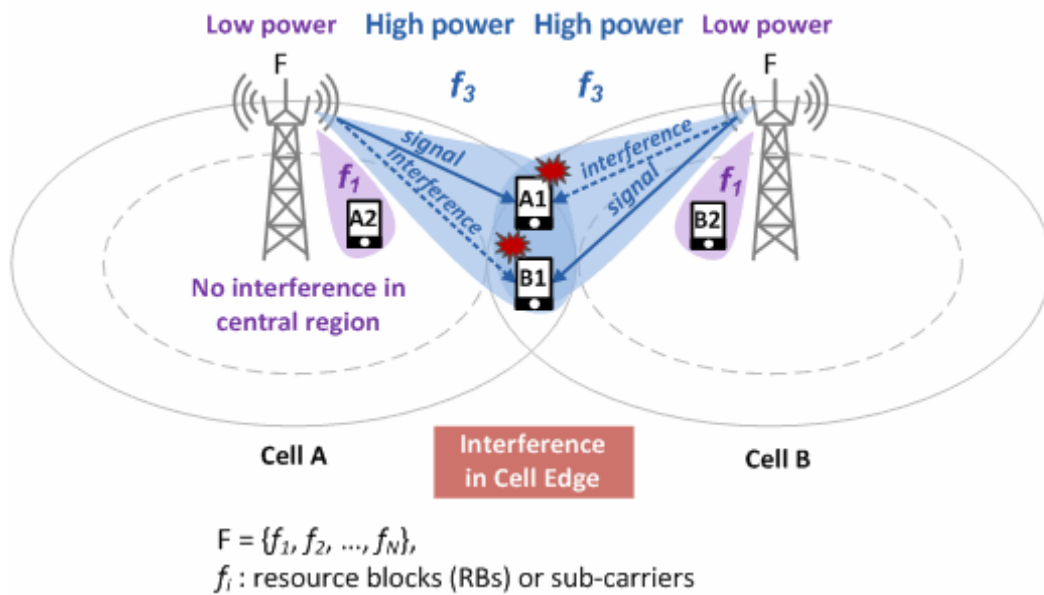

Figure 2: The interference problem in cellular networks. As the cell size decreases, inter-cell interference becomes more problematic.

## 2. Related Work

In a traditional receiver baseband system, the down-converted signal is passed through the OFDM demodulator to extract the value of subcarriers that should be categorized into one of the possible QAM symbols by using Maximum Likelihood Estimation (MLE). However, the value of the subcarriers is distorted because of the channel frequency response. By using the pilot subcarriers, the channel response can be estimated and equalized with different DSP such as Minimum-Mean Squared Error (MMSE), Zero Forcing (ZF), and Sphere Decoder (SD) algorithms.

Cell association and power control are two commonly used strategies for interference management and coordination [5]. Cell association means properly associating mobile user equipment to the serving base stations. Popular schemes include Reference Signal Received Power (RSRP)-Based Scheme [6], Bias-Based Cell Range Expansion (CRE) [7], Almost Blank Sub-frame (ABS) Ratio-Based Scheme [8]. Power control means minimizing the power (and hence minimizing the interference to other links) while keeping the desired link quality. Popular schemes include Target-SIR-tracking Power Control (TPC) [9], TPC with Gradual Removal (TPC-GR) [10], Opportunistic Power Control (OPC) [11].

There have been attempts to apply Reinforcement Learning (RL) in interference control by optimizing power transmission strategies. An RL-based decentralized power control strategy is proposed in [12], in which small cells jointly estimate the time-average performance and optimize the probability distribution for interference management in closed-access small cell networks. More recently, [13] proposed an RL-based downlink interference control scheme for ultra-dense small cell systems, in which a base station optimizes the transmit power without knowing the channel states of the neighboring cells and suppresses the inter-cell interference.

To the best of our knowledge, all the existing methods focus on finding out an optimal strategy to associate multi-tier cells and manage transmission power to control the interference. Therefore, previous works have used AI in the control layer, we are the first one to utilize AI for realtime transformation of the data.

## 3. A blind approach of interference cancellation using neural networks

We believe that the problem of interference in wideband cellular communication can be addressed using neural networks. In the training stage, the network learns the complex behavior of wideband interference using training data. The trained model then serves as a real-time signal processing module during data transmission. To address the computational load and latency issues, we will optimize the neural network for real-time operation through model compression techniques such as network pruning and coarse quantization [14].

### 3.1 Neural Network-Enhanced Receiver

Our concept for a deep learning-based receiver for next-generation wireless communication is shown in Figure 3. Instead of replacing the conventional DSP, we augment it by neural networks in the DSP pipeline to perform tasks that cannot be performed well by conventional DSP. These include wideband error compensation for the analog to digital converter (ADC) and the cancellation of inter-cell interference. In this paper, we will mainly focus on the interference cancellation part.

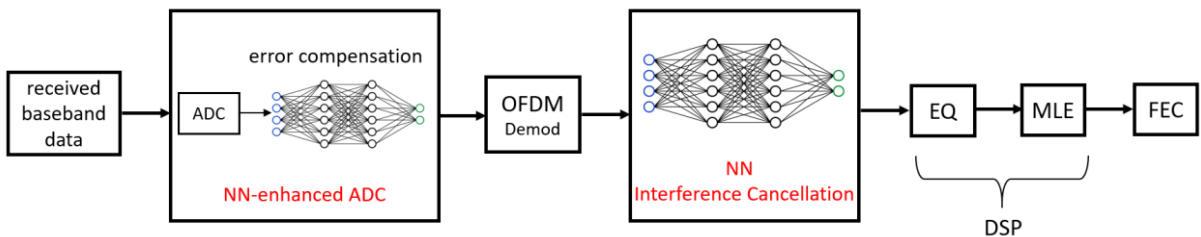

Figure 3: AI-enhanced receiver for next-generation wireless communication. OFDM: orthogonal frequency division multiplexing; EQ: channel equalization; MLE: maximum likelihood estimation; FEC: forward error correction

### 3.2 Convolutional LSTM Autoencoder

Interference from an unknown channel appears as random noise. However, in contrast with physical noise such as thermal or shot noise, the interference signal is not completely random because it belongs to a discrete QAM constellation. The subtle insight is the key that allows us to learn and remove the interference in a blind fashion. Among numerous network architectures, autoencoders have been successful for image and speech denoising [15] [16] and serve as a good starting point. For application to wireless communication, we implement the autoencoder with convolutional layers and LSTM [17] (long short term memory) layers, as shown in Figure 4. The network receives a block of symbols corrupted by interference and reconstructs the correct symbols at once. Convolutional layers perform feature extraction and LSTM layers function as an autoencoder for removing the noise-like interference. Mean Squared Error (MSE) between the recovered symbol block and ideal symbol block is used as the loss function.

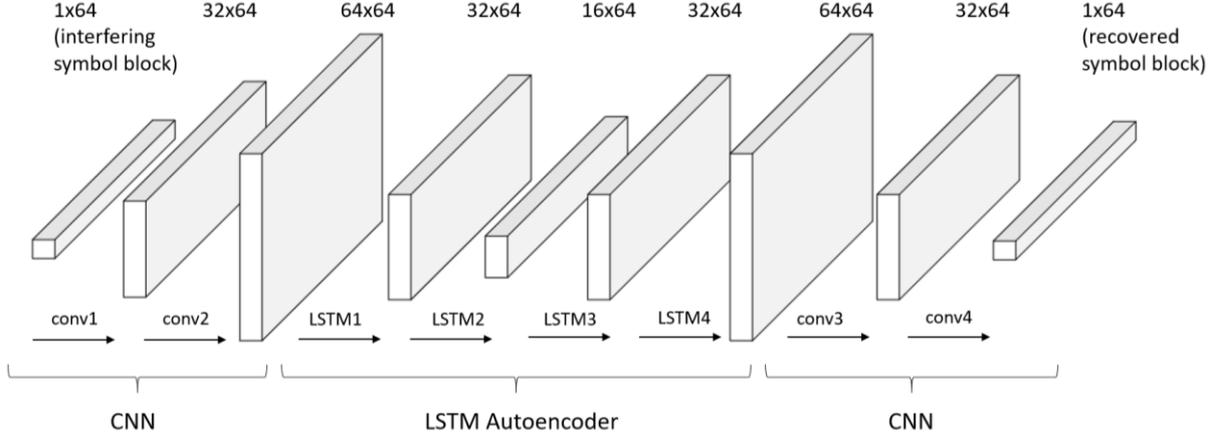

- input: interfering symbol block with length $N$ ($N = 64$): $y_1, y_2 ..., y_N$
- output: recovered symbol block with length $N$ ($N = 64$): $\hat{x}_1, \hat{x}_2 ..., \hat{x}_N$
- ground truth: ideal symbol block with length $N$ ($N = 64$): $x_1, x_2 ..., x_N$
- loss function: mean squared error between recovered symbol and ideal symbol $L = \sum_i (\hat{x}_i - x_i)^2$

Figure 4: Block diagram of our neural network for interference cancellation. The network consists of LSTM layers sandwiched between convolutional layers in an autoencoder configuration. As shown below, this architecture is very effective in blind cancellation of interference.

| | |
|---|---|
| Conv1 | Conv1d (in channel: 1, out channel:32, kernel 3x3, stride:1) + BatchNorm1d + ReLU |
| | Conv1d (in channel:32, out channel:32, kernel 3x3, stride:1) + BatchNorm1d + ReLU |
| Conv2 | Conv1d (in channel: 32, out channel:64, kernel 3x3, stride:1) + BatchNorm1d + ReLU |
| | Conv1d (in channel:64, out channel:64, kernel 3x3, stride:1) + BatchNorm1d + ReLU |
| LSTM1 | LSTM (in features: 64, hidden state features: 32, layer: 1) |
| LSTM2 | LSTM (in features: 32, hidden state features: 16, layer: 1) |
| LSTM3 | LSTM (in features: 16, hidden state features: 32, layer: 1) |
| LSTM4 | LSTM (in features: 32, hidden state features: 64, layer: 1) |
| Conv3 | Conv1d (in channel: 64, out channel:32, kernel 3x3, stride:1) + BatchNorm1d + ReLU |
| | Conv1d (in channel:32, out channel:32, kernel 3x3, stride:1) + BatchNorm1d + ReLU |
| Conv4 | Conv1d (in channel: 32, out channel:1, kernel 1x1, stride:1) |

Table 1: details of convolutional LSTM autoencoder architecture

## 4. Results

We generate 1000 radio frames in total through simulation. Each frame consists of 11 subframes, each subframe has 140 OFDM symbols and each OFDM symbol has 180 subcarriers. 500 frames are used for training, 100 frames are for validation and the rest 400 frames are for testing. The bandwidth is 3MHz with 15KHz as the subcarrier spacing and 256QAM is used as the modulation scheme. We show the results of interference cancellation in Figure 5. From the constellation map of the interfering symbols in Figure 5(a) and recovered symbols for the same frame in Figure 5(b), we can observe that the symbols suffer from severe interference and the NN does a good job in interference cancellation. We also visualize and compare the symbol error rate (SER) distribution of all the testing frames before and after interference cancellation in Figure 5(c) and can observe there is also a significant improvement on the SER. The average SER of the 400 testing frames drops from 0.37618 to 0.0003

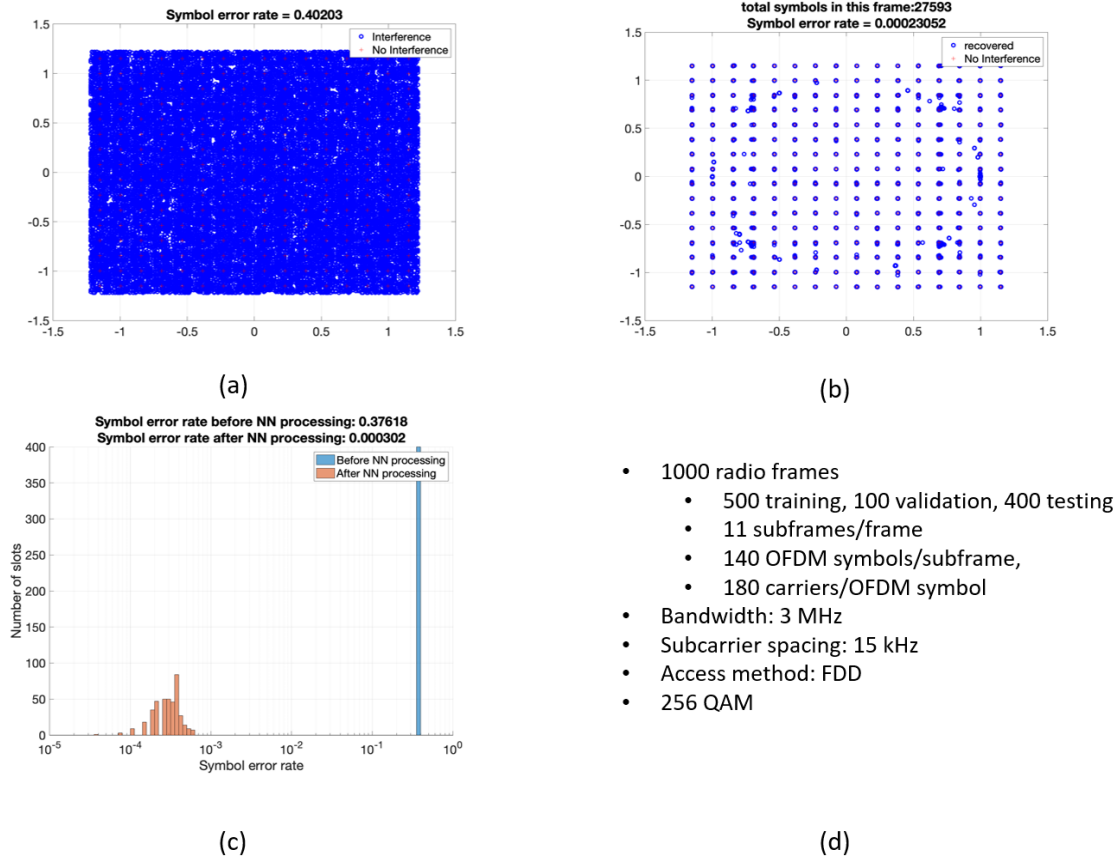

Figure 5: NN removes inter-cell interference.
(a): constellation map of interfering symbols. (b): constellation map of recovered symbols.
(c): the SER distribution of 400 testing frames w/o interference cancellation. (d): data description

## 5. Towards Hardware Implementation

Instead of floating-point computation in the software simulation, the neural network implemented in FPGA or ASIC should follow fixed-point computation rules. There has been work on the quantization of neural networks in both training and testing phase to realize efficient and accurate integer-arithmetic-only inference [18] [19] as well as the deployment of popular neural network architectures such as convolutional neural networks (CNN) and recurrent neural networks (RNN) in FPGA [20] [21], which shows the feasibility of hardware implementation for our proposed architecture.

Most of the prior work has been on classification tasks which are more tolerant of quantization errors. The problem we are discussing here is essentially a regression problem that is more sensitive to quantization errors. Nevertheless, QAM length up to 1024 should be achievable with carefully designed fixed-point arithmetic.

Furthermore, latency and power consumption are critical for applications of wireless communication. Our preliminary results have shown that NN is effective in canceling interference and crosstalk. Here we answer the question of whether our NN can be implemented in hardware with sufficiently low latency and power consumption. A Xilinx FPGA RFSoC is used to implement our proposed system. This FPGA includes 8 instances of 12-bit 4GS/s ADC. Therefore, this FPGA can be used by just simply connecting the output of the anti-aliasing filter into its analog port and do the rest of the processing on the FPGA. The whole DSP baseband processing can be implemented in the FPGA. The power consumption is directly dependent on the clock frequency used for baseband processing. The neural network processing can result in an additional 200-clock cycle to the latency of the whole chain. With an FPGA clock frequency of 200 MHz, the NN power consumption approximately results in additional power consumption of approximately 1 Watt. The power consumption will be much lower with an ASIC (instead of FPGA) and can be further reduced through model compression. The latency will be 1 microsecond which is negligible compared to the target latency of 1ms for 6G networks.

# 6. Conclusion

In this paper, we discuss a deep learning-based blind approach to remove the inter-cell interference in the physical layer for wireless communication. Our promising results show that the autoencoder architecture with the combination of convolutional layers and LSTM layers is very effective. Estimation of latency and power consumption shows the feasibility of hardware implementation in future generation communication systems. Our work indicates the direction of using neural networks in the DSP pipeline to enhance the communication system.

# References

[1] T. O'Shea and J. Hoydis, "An introduction to deep learning for the physical layer," *IEEE Transactions on Cognitive Communications and Networking,* pp. 563--575, 2017.

[2] S. Dörner, S. Cammerer, J. Hoydis and S. Ten Brink, "Deep learning based communication over the air," *IEEE Journal of Selected Topics in Signal Processing,* pp. 132--143, 2017.

[3] Z. Ghadialy, "The 3G4G Blog," [Online]. Available: https://blog.3g4g.co.uk/search/label/Picocells .

[4] DARPA, "U.S. Government System for Award Management," January 2020. [Online]. Available: https://beta.sam.gov/opp/a09982b4ddc54349a4845f106752ff50/view.

[5] E. Hossain, M. Rasti, H. Tabassum and A. Abdelnasser, "Evolution toward 5G multi-tier cellular wireless networks: An interference management perspective," *IEEE Wireless Communications,* vol. 21, pp. 118--127, 2014.

[6] J. Sangiamwong, Y. Saito, N. Miki, T. Abe, S. Nagata and Y. Okumura, "Investigation on cell selection methods associated with inter-cell interference coordination in heterogeneous networks for LTE-advanced downlink," in *17th European Wireless 2011-Sustainable Wireless Technologies*, 2011.

[7] I. Guvenc, "Capacity and fairness analysis of heterogeneous networks with range expansion and interference coordination," *IEEE Communications Letters,* pp. 1084--1087, 2011.

[8] J. Oh and Y. Han, "Cell selection for range expansion with almost blank subframe in heterogeneous networks," in *2012 IEEE 23rd International Symposium on Personal, Indoor and Mobile Radio Communications-(PIMRC)*, 2012.

[9] G. Foschini and Z. Miljanic, "A simple distributed autonomous power control algorithm and its convergence," *IEEE transactions on vehicular Technology,* pp. 641--646, 1993.

[10] M. Rasti, A. Sharafat and J. Zander, "Pareto and energy-efficient distributed power control with feasibility check in wireless networks," *IEEE Transactions on Information Theory,* pp. 245--255, 2010.

[11] K. Leung and C. Sung, "An opportunistic power control algorithm for cellular network," *IEEE/ACM Transactions on Networking,* pp. 470--478, 2006.

[12] M. Bennis, S. Perlaza, P. Blasco, Z. Han and H. Poor, "Self-organization in small cell networks: A reinforcement learning approach," *IEEE transactions on wireless communications,* pp. 3202--3212, 2013.

[13] L. Xiao, H. Zhang, Y. Xiao, X. Wan, S. Liu, L. Wang and H. Poor, "Reinforcement learning-based downlink interference control for ultra-dense small cells," *IEEE Transactions on Wireless Communications,* pp. 423--434, 2019.

[14] S. Han, H. Mao and W. Dally, "Deep compression: Compressing deep neural networks with pruning, trained quantization and huffman coding," *arXiv preprint arXiv:1510.00149,* 2015.

[15] P. Vincent, H. Larochelle, I. Lajoie, Y. Bengio, P. Manzagol and L. Bottou, "Stacked denoising autoencoders: Learning useful representations in a deep network with a local denoising criterion.," *Journal of machine learning research,* 2010.

[16] X. Lu, Y. Tsao, S. Matsuda and C. Hori, "Speech enhancement based on deep denoising autoencoder.," in *Interspeech*, 2013, pp. 436--440.

[17] S. Hochreiter and J. Schmidhuber, "Long short-term memory," *Neural computation,* vol. 9, pp. 1735--1780, 1997.

[18] R. Krishnamoorthi, "Quantizing deep convolutional networks for efficient inference: A whitepaper," *arXiv preprint arXiv:1806.08342,* 2018.

[19] B. K. S. Jacob, B. Chen, M. Zhu, M. Tang, A. Howard, H. Adam and D. Kalenichenko, "Quantization and training of neural networks for efficient integer-arithmetic-only inference," in *Proceedings of the IEEE Conference on Computer Vision and Pattern Recognition*, 2018.

[20] J. Qiu, J. Wang, S. Yao, K. Guo, B. Li, E. Zhou, J. Yu, T. Tang, N. Xu, S. Song and Y. Wang, "Going deeper with embedded fpga platform for convolutional neural network," in *Proceedings of the 2016 ACM/SIGDA International Symposium on Field-Programmable Gate Arrays*, 2016.

[21] Y. Guan, Z. Yuan, G. Sun and J. Cong, "FPGA-based accelerator for long short-term memory recurrent neural networks," in *2017 22nd Asia and South Pacific Design Automation Conference (ASP-DAC)*, 2017.